\documentclass[doublespacing]{elsart}
\begin{document}
\begin{frontmatter}
\title{Assemblies of gold icosahedra}
\author{G. Bilalbegovi\'c\thanksref{postadr}}
\ead{goranka.bilalbegovic@zg.htnet.hr}
\thanks[postadr]{Fax: +1-775-414-7947; Tel:+385-98-480-921; Postal address (home): Grada Vukovara 43A, 10000 Zagreb, Croatia}
\address{Department of Physics, University of Rijeka, Omladinska 14, 51000 Rijeka, Croatia}

\begin{abstract}
Low-dimensional free-standing aggregates  of bare gold clusters are studied by
the molecular dynamics simulation. Icosahedra  of $55$ and $147$ atoms are
equilibrated at $T=300$ K. Then, their one- and two-dimensional assemblies are investigated.
It is found that icosahedra  do not coalescence into large drops, but stable amorphous nanostructures are formed:
nanowires for one-dimensional and nanofilms for two-dimensional assemblies. The high-temperature stability of these nanostructures is also investigated. 
\end{abstract}

\end{frontmatter}

\section{Introduction}
Gold is very interesting for applications in nanotechnology
\cite{Wyrwa,Reuter,Whetten,Luedtke,Mohamed,Diao,OShea,Hutchinson,Fullam,Hoeppener,Sun}.
All bare low-index surfaces of gold exhibit reconstructions
and act as very good substrates for inorganic, organic and biological thin films.
Nanoparticles of gold show interesting optical and electronic properties. Gold
clusters arranged in one and two dimensions are important for applications, most notably
as quantum-dot arrays.

Isolated gold clusters are extensively studied by experimental and theoretical methods 
\cite{Martin,Andreoni,Garzon}.
Various amorphous, crystalline and quasicrystalline structural forms were found. Ordered high symmetry structures are
Mackay icosahedral and fcc cuboctahedral configurations.
A producing of cluster-assembled materials is nowadays very important. The discovery of $C_{60}$, other fullerenes,
and fullerite indicates the route towards synthesis of new materials consisting of stable clusters.
Gold icosahedra are stable and therefore, they can be used as a building element of assembled materials.

To stabilize aggregates of gold clusters various methods are used, for example, the Langmuir-Blodgett technique,
addition of functionalized polymers, or DNA \cite{Wyrwa,Reuter,Whetten,Luedtke}. Gold nanoparticles and nanorods are also successfully synthesized by the electrochemical method \cite{Mohamed}. Recently, gold nanoparticle films were prepared by the surface vertical deposition method on a wettable substrate \cite{Diao}.
It is especially difficult to make and investigate short one-dimensional arrangements of gold nanoparticles.
Such structures were obtained for $Au_{55}(PPh_3)_{12}Cl_6$ clusters \cite{Wyrwa}.
These $55$ atoms gold clusters surrounded by ligand shells were positioned along bundles of polymers.
Networks of polymer chains decorated by gold clusters were deposited on the substrates using the Langmuir-Blodgett technique \cite{Reuter}.
It is possible to prepare these networks with different densities.
Nanostructured Au cellular networks were produced by spin coating
a solution of thiol-passivated gold nanoparticles on the silicon substrate \cite{OShea}.  It was found that
annealing at $(500-600)$ K removes thiol surfactants, but the cellular structure of bare gold nanoparticles was preserved.
One-dimensional organizations of gold clusters were also made by decoration of various templates, for example carbon nanotubes \cite{Hutchinson,Fullam,Hoeppener}.
Gold nanoboxes with a truncated cubic shape where obtained using silver nanocubes as templates \cite{Sun}.
These bare gold nanoboxes self-assembled into a close-packed two-dimensional array.

In this work the Molecular Dynamics (MD) simulation results for ordered low-dimensional aggregation of small gold icosahedra are presented. It is shown that, although a coalescence of atoms occurs at the touching point of neighboring clusters, stable one- and two-dimensional aggregates are formed at room temperatures. Assembled nanostructures 
are stable up to at least $800$ K. 
These results are important for understanding of processes and development of new techniques in preparation of gold nanowires and thin films along the route described in Refs.
\cite{Wyrwa,Reuter,Whetten,Luedtke,Mohamed,Diao,OShea,Hutchinson,Fullam,Hoeppener,Sun}.
In the following a computational method is described in Sec. II. The results of simulation and
a discussion are presented in Sec. III. A summary and conclusions are given in Sec. IV.

\section{Computational Method}

I have used a computer simulation to elucidate the stability of low-dimensional arrangements of gold icosahedra.
The classical MD method is carried out \cite{Frenkel}.
This method is very suitable for modeling of large systems on long time scales.
It is not yet feasible to study these large scale problems within ab-initio methods.
It is well known that the pair potentials in classical simulations do not give a correct description of metal properties  \cite{Daw}.
Instead, many-body potentials of the embedded-atom type are successfully applied for metals.
In this work the ``glue'' realization of embedded atom potentials
is used to describe the interparticle interactions \cite{Furio}.
This potential is very well tested
and known to give good results for many bulk, surface and nanoparticle properties
\cite{Furio,Andreoni,PRB,SSC,JPCM,Tosatti,Baolin,Vac}. In contrast to several other potentials
used in classical simulations, this one reproduces different reconstructions on all low-index gold surfaces
\cite{Furio}. Previous simulations of gold clusters using the same potential confirmed several results obtained experimentally \cite{Andreoni}.
Therefore, it is expected that simulated aggregations of gold icosahedra realistically model natural
nanostructures.
The temperature of the systems
is controlled by rescaling particle velocities. The time step used is $7.14 \times 10^{-15}$ s.

The clusters are prepared  as ideal icosahedra of $55$ and $147$ atoms at $T=0$ K. After a procedure of
annealing and quenching, the clusters are equilibrated  for $10^5$ time steps at $T=300$ K. Several fully equilibrated  clusters are then placed in contact with one another and aligned in one or two dimensions.
The contact of icosahedra in one dimension is shown in Figs. 1 (a) and (c).
Two, five and ten icosahedra are positioned in one dimensions. The arrangements of $5 \times 5$
clusters are made in two dimensions.
Resulting structures are equilibrated for additional $10^6$ time steps, i.e., for $7.1$ ns. 
The assembled nanostructures are also heated up to $1400$ K, in steps of $100$ K. They were equilibrated for $10^6$ 
time steps at each temperature. The experimental bulk melting point of gold is $1336$ K, whereas for the potential used 
in this simulation the bulk melting temperature was fitted  to $1357$ K
\cite{Furio}.

\section{Results and discussion}

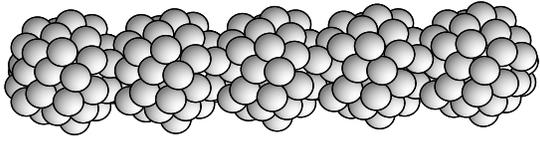
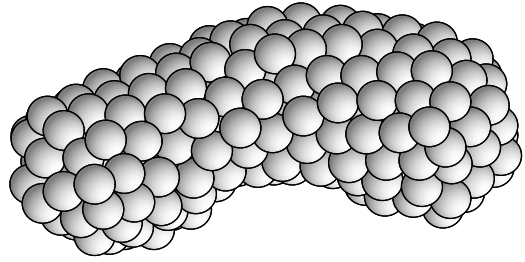
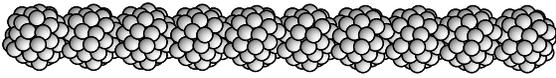
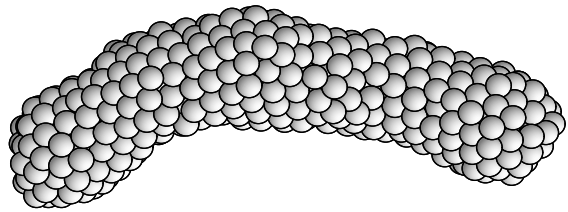
\begin{figure}
\caption{Atomic positions at $300$ K for: (a) five assembled $55$ icosahedra,
(b) the nanostructure shown in (a) after $10^6$ steps of time evolution,
(c) ten assembled $55$ icosahedra,
(d) the structure shown in (c) after $10^6$ time steps of evolution.}
\label{fig1}
\end{figure}

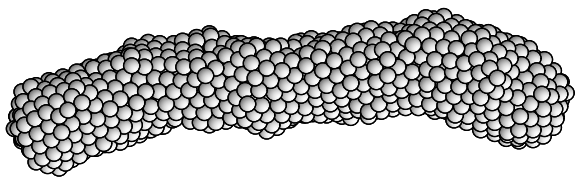
\begin{figure}
\caption{Atomic positions at $300$ K for the equilibrated nanostructures assembled from:
(a) five $147$ icosahedra,
(b) ten $147$ icosahedra.}
\label{fig2}
\end{figure}

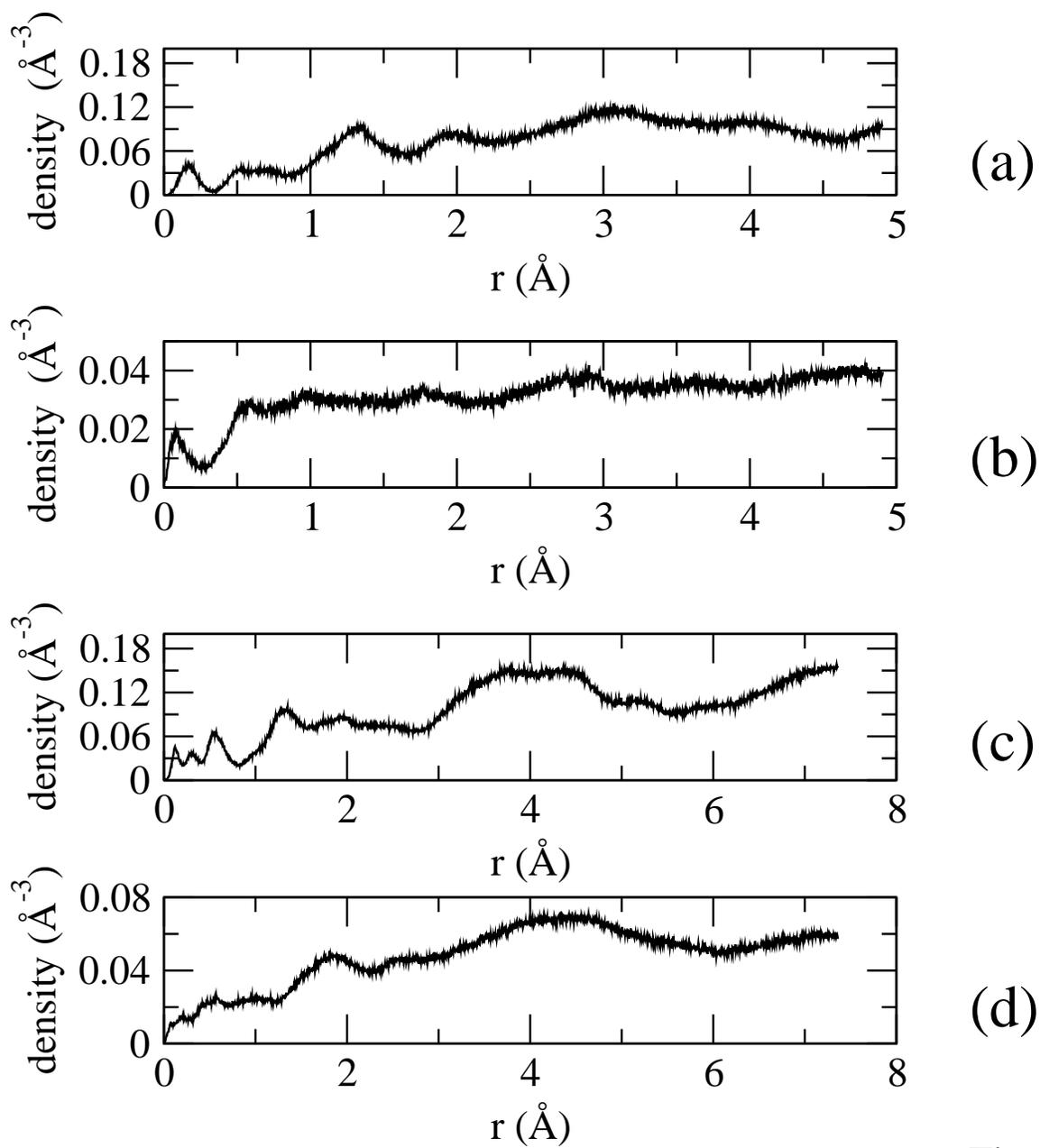
\begin{figure}
\caption{The radial density plots at $T=300$ K for the equilibrated nanowires assembled from:
(a) five $55$,
(b) ten $55$,
(c) five $147$,
(d) ten $147$ atoms icosahedra.}
\label{fig3}
\end{figure}

An assembly of only two icosahedra evolves into an elongated, potato-shaped nanoparticle.
Figure \ref{fig1} presents one-dimensional assemblies of five and ten icosahedra
consisting of $55$ atoms.
Wormlike nanowires are obtained, as shown in Fig. \ref{fig1} (b) and (d).
Figure \ref{fig2} shows one-dimensional assemblies of $147$ atoms icosahedra after $10^6$ time steps of evolution.
Nanowires in Fig. \ref{fig2} are less curved than one-dimensional aggregates of $55$ icosahedra.
The radial density plots for nanowires shown in Figs. \ref{fig1} (b), (d) and Fig. \ref{fig2} 
are presented in Fig. \ref{fig3}.
A radial ordering does not exist. This is in contrast with radial structures of
gold nanowires studied previously in MD simulations \cite{PRB,SSC,JPCM,Tosatti,Baolin,Vac}.
For example, the starting point in Refs. \cite{PRB,SSC,JPCM,Vac}
was a cylindrical nanostructure carved from regular fcc lattice of bulk gold. In MD simulations
these structures evolved to multi-shell nanowires similar to multi-wall carbon nanotubes.
Multi-shell gold nanowires were also observed by  high-resolution
transmission electron microscopy \cite{Kondo}. 

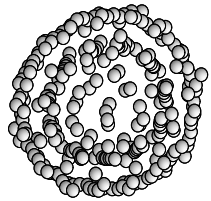
\begin{figure}
\caption{Atomic positions for a nanowire shown in Fig. \ref{fig1} (b) after the heating and equilibration procedure at
$800$ K:
(a) side view,
(b) cross-section.}
\label{fig4}
\end{figure}

In previous simulations it was found that the best multi-wall structures exist at temperatures
$(800-900)$ K \cite{SSC}. Nanostructures produced by self-assembling of icosahedra and simulated in this work are
also studied at higher temperatures.  As in previous simulations \cite{SSC,Vac} the mean-square displacements and diffusion coefficients plots show that the melting starts above $900$ K. 
Figure \ref{fig4} presents the nanostructure assembled from five $55$ icosahedra at $800$ K. The side view 
in Fig. \ref{fig4} (a) shows that this structure is much more straight than at $300$ K (see Fig. \ref{fig1} (b)). 
The cross-section in Fig. \ref{fig4} (b) shows multi-shells at an initial stage of development. 
The annealing and quenching procedure used for a gold nanowire preparation in previous simulations 
\cite{PRB,SSC,JPCM,Vac} produces much better shells. 
The structure assembled from ten $55$ atoms icosahedra (see Fig. \ref{fig1} (d)) also starts to
order above the room temperature. However, the kink develops at $600$ K, and the structure at $900$ K
evolves to a spherical solid drop. The structure assembled from five $147$ icosahedra (see Fig. \ref{fig2} (a)) has a nanowire morphology at $(300-800)$ K and evolves to a spherical solid shape at $900$ K. 
The nanowire made from ten $147$ icosahedra (shown in Fig. \ref{fig2} (b))
evolves first to irregular (``potato'') morphology at $900$ K, and
then transforms to the spherical solid shape at $1100$ K.

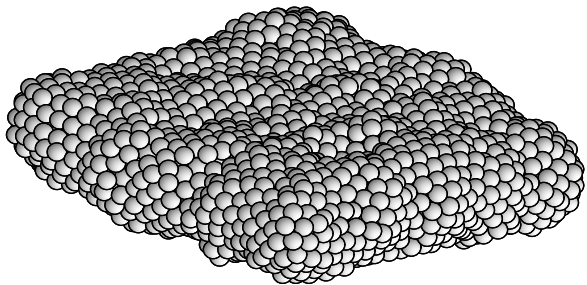
\begin{figure}
\caption{MD snapshots of atomic positions at $300$ K for the equilibrated nanofilms assembled from:
(a) $55$ atom icosahedra,
(b) $147$ atom icosahedra.}
\label{fig5}
\end{figure}

Figure \ref{fig5} shows two-dimensional assemblies after $10^6$ time steps of evolution at $300$ K. 
Stable thin nanofilms are formed. Their surfaces are  (111) oriented with many local defects.
Transition of gold (001) nanofilms to the (111) orientation induced by local thinning
was observed by the electron microscope \cite{Ru}.
It was found there that the (111) film expands laterally and then breaks.
The smallest size of nanofilms studied in Ref. \cite{Ru} was $(2 \times 2)$ nm$^2$, 
and the largest one was $(24 \times 14)$ nm$^2$.
The thickness of the (111) film just after transition from (100) was about $2$ nm.
Nanofilms studied here are $(4.9 \times 4.9)$ nm$^2$ and $(7.4 \times 7.4)$ nm$^2$. They are both
thinner than $2$ nm. The thickness of the film made of $55$ atom icosahedra
is $\sim 1$ nm. The film assembled from $147$ atom clusters  is $1.5$ nm thick.
For both nanofilms this morphology is stable up to $1000$ K. At higher temperatures nanofilms transform to 
the spherical shape.

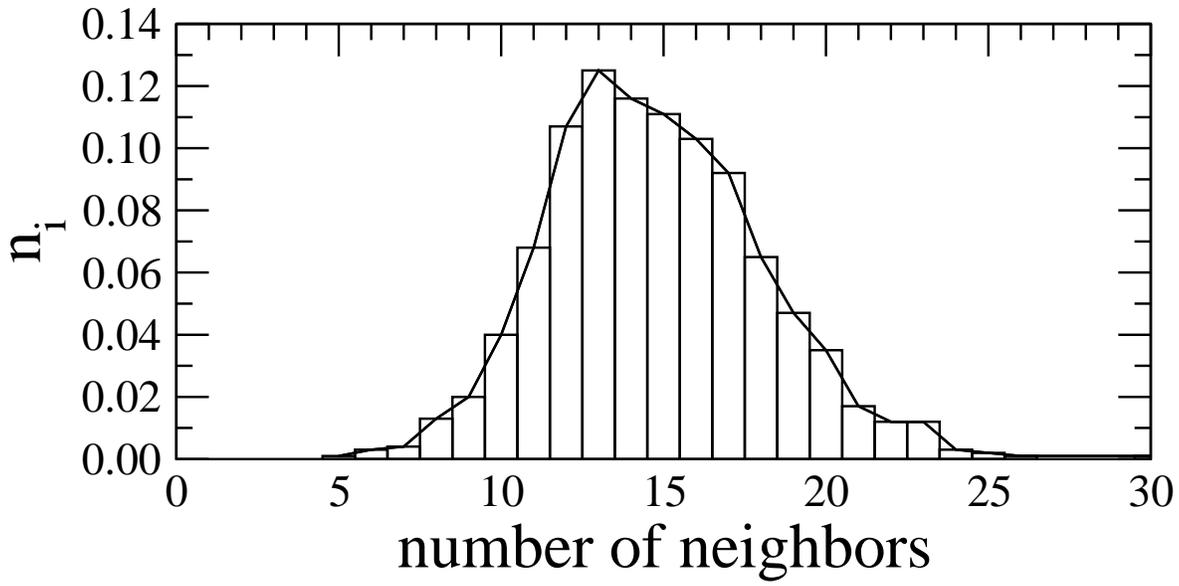
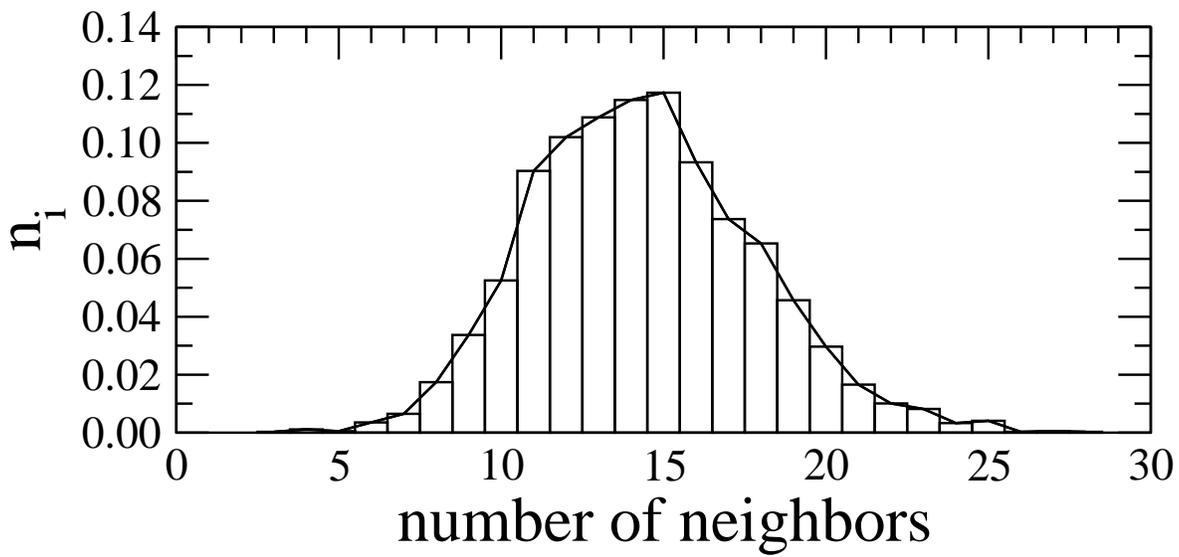
\begin{figure}
\caption{The fractional concentrations for the number of faces of the Voronoi polyhedra in nanofilms 
shown in Fig. \ref{fig5} and 
assembled from:
(a) $55$ atom icosahedra,
(b) $147$ atom icosahedra. The bars show the numbers, whereas the solid line is the envelope.}
\label{fig6}
\end{figure}

A structural analysis of nanofilms shown in Fig. \ref{fig5}
is carried out via the Voronoi construction \cite{Collins}. This construction is
a topological generalization of the Wigner-Seitz one in solid state physics. The Voronoi polyhedron for a given atom is defined as the region consisting of all points nearer to it than to any other atom. The Voronoi construction represents local atomic configurations and disorder 
for a given structure.
Figure \ref{fig6} presents the fractional concentrations for the 
number of neighbors calculated for the atoms in nanofilms shown in Fig. \ref{fig5}.
The particles are not assumed to be neighbors beyond the cutoff of $5$ \AA.
The mean number of the Voronoi
neighbors $<n>$ is $14.75$ for the distribution shown in Fig. \ref{fig6} (a),
and $14.40$ for Fig. \ref{fig6} (b). It is known that $<n>$ is $6$ in two dimensions, and $14$ in three dimensions and for regular close packing \cite{Collins}. In a MD simulation Rahman found $15.67$ for the randomly distributed atoms of
ideal gas, $14.45$ for the liquid argon, and $14.26$ for the solid argon \cite{Rahman}. The maximum of distribution
for the number of faces is $13$ for Fig. \ref{fig6} (a), and $15$ for Fig. \ref{fig6} (b). 
The broadness of the distributions in Fig. \ref{fig6} shows the presence of topologically defective polygons where
the number of faces is not $14$.

It is known that bare metal clusters deposited on a substrate diffuse and often show a tendency towards coalescence and sintering. The processes of arrangements of gold icosahedra simulated here do not produce large spherical drops
at $300$ K, and up to $800$ K. 
This type of  coalescence is inhibited by a limited diffusion at lower temperatures and the presence of facets on initial clusters. Similar results were found in a previous MD simulation for aggregation of two solid gold nanoparticles \cite{Lewis}. This was explained as a result of the presence of very small facets on spherical solid clusters. The initial clusters studied here are much more regular and facetted than in Ref. \cite{Lewis}, where spherical $(135-3997)$ atom nanoparticles were carved from a block of fcc bulk gold lattice. Large solid facets on icosahedra studied here prevent strong interdiffusion and sintering into large solid drops. 
\bigskip

\section{Conclusions}

Molecular dynamics simulation of one- and two-dimensional aggregations of free bare gold icosahedra has been carried out.
Assembling of bare gold clusters may be used as a method for designing and fabrication of nanowires and nanofilms.
It is possible to prepare gold icosahedra in solutions, or in the gas phase using supersonic beams.
Similar arrangements of gold clusters as those simulated here, may be prepared on the substrates
by controlled deposition, or using STM and AFM tips.

\ack{This work has been carried under the HR-MZT project 0119255 ``Dynamical
Properties and Spectroscopy of Surfaces and Nanostructures''.}

\clearpage

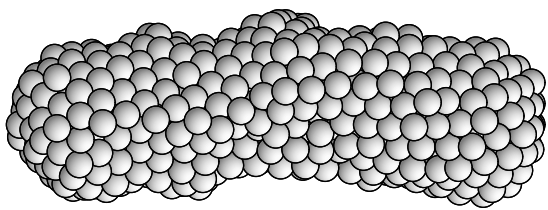
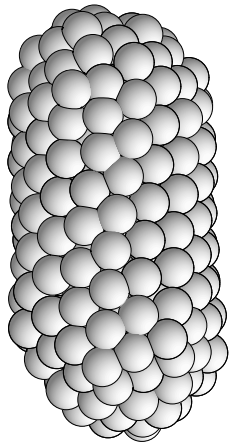
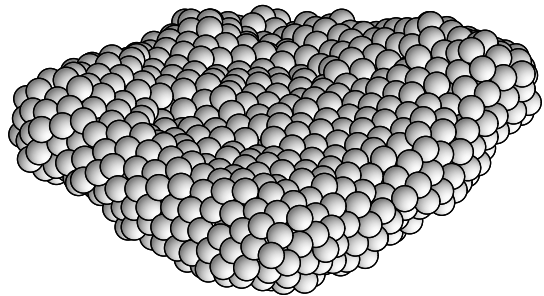
\end{document}